# Ultrasensitive Detection Enabled by Nonlinear Magnetization of Nanomagnetic Labels

M. P. Nikitin[1,2,3], A. V. Orlov[3], I. L. Sokolov[2], A. A. Minakov[3], P. I. Nikitin[3], J. Ding[1], S. D. Bader[1], E. A. Rozhkova[4], V. Novosad[1,5].

[1] Materials Science Division, Argonne National Laboratory, Argonne, IL 60439, USA.

[2] Moscow Institute of Physics and Technology, Dolgoprudny, Moscow Region 141700, Russia.

[3] Prokhorov General Physics Institute, Moscow 119991, Russia.

[4] Center for Nanoscale Materials, Argonne National Laboratory, Argonne, IL 60439, USA.

[5] National University of Science and Technology (MISiS), Moscow, 119049, Russia.

**ABSTRACT**: Geometrically confined magnetic particles due to their unique response to external magnetic fields find a variety of applications, including magnetic guidance, heat and drug delivery, magneto-mechanical actuation, and contrast enhancement. Highly sensitive detection and imaging techniques based on *nonlinear properties* of nanomagnets were recently proposed as innovative strong-translational potential methods applicable in complex, often opaque, biological systems. Here we report on significant enhancement of the detection capability using optical-lithography-defined, ferromagnetic iron-nickel alloy disk-shaped particles. We show that an irreversible transition between a strongly non-collinear (vortex) and single domain states, driven by an alternating magnetic field translates into a nonlinear magnetic response that enables ultrasensitive detection of these particles. The record sensitivity of ~ $3.5 \times 10^{-9}$ emu, which is equivalent to ~39 pg of magnetic material is demonstrated at room temperature for arrays of patterned disks. We also show that unbound disks re-suspended in aqueous buffer can be successfully detected and quantified in real-time when administrated into a live animal allowing for tracing their biodistribution. Use of nanoscale ferromagnetic particles with engineered nonlinear properties opens prospects for further enhancing sensitivity, scalability and tunability of noise-free magnetic tag detection in high-background environments for various applications spanning from biosensing and medical imaging to anti-counterfeiting technologies.

Nanoparticles are becoming prominent in life sciences for numerous applications which span from intelligent sensing and theranostics [1-3] to controlling cell function [4] and futuristic nanorobotics devices.[5] Magnetic particles (MP) have been attractive for advancing medical technologies due to their unique functionality, possibility to control their magnetization, location and temperature remotely by the external fields, as well as their relatively low cost and environmental safety. MPs and their hybrids with (bio)organic materials were proposed for drug delivery, cell sorting, hyperthermia, and non-invasive medical imaging.[6-10] When MPs are interfaced with a living cell they can be used to manipulate within biological machinery and alter biological functions at various levels of biosystem complexity, from organelle and single cell to neural networks and model vertebrate organisms.[11-15]

The majority of MP-based nano-labels are utilized in magnetic resonance imaging (MRI) modality where they enhance contrast due to spatial interaction of their magnetic moments with local water molecules.[16-17] Unlike MRI, the newer magnetic particle imaging (MPI) [18] and magnetic particle quantification (MPQ) [19-21] techniques rely upon direct interaction of the MP with an alternating magnetic field and exploit the fact that the field-driven magnetic response has nonlinear features, in particular closer to the saturation region. Thus, by applying an oscillating magnetic field and registering the signal at higher harmonics of the excitation source it is possible to detect the magnetic particles remotely, often with the sensitivity that surpasses the traditional methods. The output signal correlates with the nonlinearities in M(H) dependence and varies with the magnitude of static magnetic field. With these techniques, MP labels can be directly detected in biological environments, for example, visualized in real time [22-23] *in vivo* within opaque organisms, [8, 18, 24-25] or precisely counted with extraordinary wide linear dynamic range (seven orders of magnitude) in three-dimensional (3D) format in complex biological fluids[19-21, 26-27] as well as in intelligent nanorobotic devices[5], with no obscuring background signal from endogenous diamagnetic or paramagnetic materials and biomolecules. This low-noise method is based on the nonlinearity of the magnetization of ferromagnetic material in an oscillating magnetic field. State-of-the-art sensitivity was reported to reach ~0.33-0.4 ng of $Fe_3O_4$ nanoparticles in 200 μL sample volume via portable MPQ detectors,[26, 28] which is on the level of γ-radioactive techniques,[25] and exceeds the sensitivity of superconducting quantum interference device (SQUID) detected magnetic relaxometry.[29-30] Along with advancing detection systems, further improvement of the MPQ and MPI techniques relies greatly on the development of magnetic labels with tailored properties optimized for stronger signals. In particular, nanomaterials with high nonlinearity of magnetization in weak magnetic fields could potentially improve detectability limits of magnetic tags in small animals using existing MPI scanners, [24] further allowing for magnetic detection in deep organs of human body, where penetrating magnetic fields from external sources should be able to magnetize a label.

**Results and Discussion**

In the present work, we tested the applicability of optical-lithography-defined ferromagnetic $Fe_{19}Ni_{81}$ (permalloy) microdisks as MPQ magnetic labels. These materials are known to possess a vortex ground state characterized by an in-plane continuously curling spin configuration with zero net in-plane magnetization in the absence of a magnetic field.[31-34] Zero remanence eliminates dipolar field-driven agglomeration when the disks are interfaced with a biological system. This essential feature, along with high value of the magnetization of saturation, allows for efficient remote detection, sensing and manipulation of the disks within biological machinery.

Top-down fabrication of the magnetic structures relies upon physical techniques including optical, nanoimprint or electron-beam lithography combined with physical vapor deposition that provides vast opportunities for production of mono-dispersed structures of tailored dimensions, composition and magnetic properties. Due to their unique properties, permalloy magnetic disks were utilized as mediators for initiating cancer cell destruction *via* magneto-mechanical actuation of a membrane in weak, oscillating magnetic fields[3-4, 12], while their nonlinear properties remained practically not explored. Here we show how an irreversible *ac* magnetic field-driven transition from a strongly non-collinear (vortex) to a collinear (single-domain) state of this nanomaterial translates into a strong nonlinearity of the magnetization process that gives rise to ultrasensitive MPQ detection. Unprecedented sensitivity of ~ $3.5 \cdot 10^{-9}$ emu, equivalent to ~39 pg of the magnetic material is demonstrated.

We will first describe the results obtained for magnetic disks patterned on silicon wafers, and then for *in vivo* experiments with unbound disks. A representative scanning electron microscope (SEM) image for a 20-nm thick, 1.5 µm in diameter dot array sample is shown in Fig. 1. The in-plane hysteresis loops are typical for a vortex nucleation, displacement and annihilation magnetization reversal process, as seen in Fig. 2a,b. At remanence, the samples possess no net magnetic moment, which is indicative of the spin vortex (SV) state.[35-36] Upon application of a magnetic field, the vortex center is displaced, and the sample gradually becomes magnetized. This process corresponds to the linear (reversible) part of the hysteresis loop. When the magnetic field reaches a critical value, $H_s$, the vortex annihilates and the sample becomes uniformly magnetized (a single-domain (SD) state) as schematically depicted in Fig. 2c. Conversely, under decreasing magnetic field the SD collapses into SV state, which is accompanied by an abrupt drop in the macroscopic magnetization. Both descending and ascending branches of the hysteresis loop coincide in small fields, when the vortex core is located close to the center of the disk.

No hysteresis is observed when the magnetic field is applied perpendicular to the disk plane (the vortex core switching in the perpendicular fields [31, 37] involves very small volume of the sample, and can be ignored). In this case, the magnetization process corresponds to reversible canting of magnetic moments out of the disk plane. The out-of-plane saturation field is significantly larger than its in-plane counterpart, thus a linear $M(H)$ dependence is observed in a broad field range (see Fig. 2b inset). The magneto-crystalline anisotropy of permalloy is small in comparison with the magnetostatic demagnetization field, $H_D$, and therefore can be neglected. Subsequently, the in-plane saturation field $H_S$ (the field required for vortex state annihilation) is comparable to $H_D = N_{\parallel} \cdot M_S$, where $N_{\parallel}$ is the corresponding demagnetizing factor. It should be noted that both the vortex nucleation and annihilation fields increase with increasing disk geometric aspect ratio, and can be calculated analytically using the "rigid vortex" model.[35]

The magnetic response of the array of permalloy disks was measured using two different MPQ devices. The MPQ detection is based the nonlinear magnetic response of the sample subjected to a two-component alternating magnetic field with amplitudes $H_1$, and $H_2$ and frequencies $f_1$ and $f_2$, correspondingly, with signal detection at a frequency $f = mf_1 \pm nf_2$ (where $m$, and $n$ are integers, one of them can be zero). The first set of experiments was performed at $H_1 = 144 \pm 10$ Oe, $H_2 = 56 \pm 5$ Oe, and $f_1 = 154$ Hz, $f_2 = 150$ kHz. The signal of this "MPQ-1" detector [26] was measured at combination frequencies $f_2 \pm 2f_1$. This detector has the limit of detection (0.4 ng) of the most "responsive" commercially available magnetic beads (196-nm Estapor®, Merck Chimie SAS, France) dispersed in 200 µL volume, and a linear dynamic range of >7 orders of magnitude. The maximum field amplitude $H_{max} = H_1 + H_2 = 200 \pm 11$ Oe of the MPQ-1 device is higher than the saturation fields for all tested disks array samples.

During the measurements, the $3 \times 3$ mm² sample was slowly translated through the MPQ device at a speed of ~150 microns per second. Figure 3a summarizes the MPQ sensor output as a function of the sample's lateral position obtained for different disk arrays parallel to the direction of the *ac* magnetic field. The signal changes continuously as the sample is displaced and reaches a maximum value when the sample is located in the center of the measurement cell. The signal increases with the volume of the disks (*e.g.,* the disk thickness). The highest signal-to-noise ratio (SNR) in these measurements was achieved for 30-nm-thick disks. The noise level was determined using a bare wafer of the same size, see inset in Fig. 3a. Specifically, as few as 87 particles which amounts to 39.4 pg of permalloy or a magnetic moment of $3.5 \times 10^{-9}$ emu ($3.5 \times 10^{-12}$ A·m$^2$) could be readily detected against the background noise. This limit of detection is at least an order-of-magnitude better than data published thus far for conventional MPQ detection that uses commercial superparamagnetic particles.[26, 28] Moreover, the sensitivity can surpass the impressive detection performance of SQUID magnetometers designed for extended objects and room-temperature applications.[29-30, 38] The measurement error, when registering inherently weak signals, is mainly due to the presence of the background electronic noise, with the peak-to-peak amplitude of ~20 r.u. (the insert in Figure 3a), whereas the magnetic signal from the sample exceeds the above value for ~ 5,000 times. It should be noted that in our measurements we averaged the signal by collecting one data point collected per second. The ability to measure small signals quickly and with high fidelity is a characteristic feature of the MPQ technique. It is especially valuable for a wide range of life science applications where a real-time monitoring of the rapidly evolving processes is crucial. The above errors, and, as a consequence, SNR, can be further improved by using a longer integration time, it will cause a deterioration in temporal resolution.

Interestingly, the disks demonstrated remarkable dependence of the MPQ signal on the angular orientation to external magnetic field. This feature can be valuable for the determination of the local rheology. Thus, subsequently the same samples were positioned in the center of the MPQ-1 sensor at different angles $\varphi$ with respect to the direction of the alternating magnetic field. Figure 3b compares the angular dependences of the signal $S(h, \varphi)$ obtained for different disk arrays. First, we find that while the signal increases with the disk thickness, there is no significant angular variation when the angle $\varphi$ remains small. With misalignment increase between the applied magnetic field and the disk plane, the signal gradually declines and approaches to minimum as $\varphi \to 90°$ for all samples. The position of the maximum slope $\partial S(h, \varphi)/\partial \varphi$ shifts to smaller angles for thicker disks (see Fig. 3c) because of the increase of their demagnetizing field $H_D$. The latter is illustrated in the plots of $S(h, \varphi)$ normalized to the $S(h, \varphi=0)$, Supplementary Fig. S1. The strongest nonlinearity of M(H) exists only if the vortex nucleation and annihilation processes are involved (see Supplementary section on analysis of hysteresis loops in alternating magnetic fields, and Fig. S2). Thus, the signal $S(h, \varphi)$ attains its maximum, when the amplitude of the in-plane component of the applied magnetic field becomes comparable to (or exceeds) the saturation field $H_s$. This explains why for some angles $S(h, \varphi)$ could be stronger for the samples that contain less magnetic material (*e.g.,* thinner disks).

To further clarify this, we performed additional measurements with another "MPQ-2" detector [28] optimized to produce smaller magnetic fields. Here, $H_1$ and $H_2$ were equal to 64 ± 6 Oe, and 33 ± 3 Oe, with $f_1 = 702$ Hz, and $f_2 = 87$ kHz, respectively. The field amplitude $H_{max} = H_1 + H_2 = 97 ± 7$ Oe is about half of that in the previous set of experiments, and is substantially less than the saturation fields for the 40- and 30-nm disks. Therefore, it is not surprising that the latter two samples show nearly no signal for the whole range of angles, see Fig. 3d. On the other hand, the 20 nm-disk array generated a strong MPQ response since $H_{max}$ was close to $H_S$ for this sample (see the hysteresis loop in Fig. 2). A key benefit of disks with large geometric aspect ratio *D/h* is that they can be detected utilizing relatively weak magnetic fields. This may become especially beneficial when imaging larger, clinically-relevant, objects. Interestingly, we find that angular dependence of the MPQ-signal is characteristic of the disk thickness (for given disk diameter). Thus, the signals from disks with different geometries should, in principle, be distinguishable by analyzing the $\partial S(h, \varphi)/\partial \varphi$ derivatives.

To validate our initial findings in a biological system, we performed a series of *in vivo* studies. For this purpose, the 1.5-µm-diameter disks were lifted off the wafer and re-suspended in phosphate buffered saline (PBS) resulting in a concentration of ~2 × 10$^9$ disks per 1 ml, following the procedures reported elsewhere [12]. The primary goal of this experiment was to assess the detectability of the disks circulating in the blood flow of mice in real-time using the MPQ technique, as well as to understand how the detector signals can be modulated by affected on disk's magnetization or orientation inside the organism by external quasistatic magnetic fields. Thus, the *in-vivo* experimental setup was configured to include an MPQ detector coil generating the excitation (*ac*) field, and a larger coil that provides an external bias (*dc*) field, see photograph shown in the Fig. 4. The magnetic disks suspended in PBS solution were injected retro-orbitally (that is via the ophthalmic venous sinus) [39] into the blood flow of the animal. The anesthetized animal was placed inside the MPQ pick-up coil, which quantified the magnetic content passing through the animal's tail veins and arteries.

First, we studied the blood circulation dynamics of the disks without application of an external *dc* magnetic field. Figure 4a shows that right after injection the magnetic signal quickly rises as the disks reach the tail blood vessels to the maximum concentration, and then exponentially declines as the disks get cleared by the reticulo-endothelial system. The disks were cleared from the blood stream within <10 min. Noteworthy, no specific biofunctionalization of the particle's surface was undertaken at this stage in order to allow for a better understanding of the magnetic disk's intrinsic performance and passive distribution *in vivo*.

During the next experiment an external *dc* magnetic field of only 15 Oe was applied in addition to the to the detector's *ac* field one minute after the disks injection. As presented in Fig 4b, while the kinetics of the particles' clearance remains similar to that in the previous experiment, the MPQ signal remarkably increased almost two-fold upon application of such a small magnetic field, and then abruptly decreased upon turning the field off. Both processes can be fitted with exponents having the same decay constant, which suggest that the magnetic particle removal rate is not affected by spatially uniform *dc* magnetic fields. In sharp contrast, the commercial ~1.0 μm spherical magnetic particles, Dynabeads®, do not demonstrate any signal enhancement in response to the application of an external magnetic field, Fig. 4c. The fact that the signal can be enhanced for the disks by a proper alignment with respect to the ac probing field but not for spherical beads is not surprising as the magnetic properties of spherical particles are expected to be isotropic. Such responsiveness of the MPQ-signal produced by the disks readily allows discrimination of the vortex labels from conventional isotropic ones, thereby providing an opportunity to multiplex magnetic labels for various *in vivo* applications and *in vitro* assays. Additionally, such uncommon behavior opens up exciting prospects for further improvement of MPQ sensitivity and noise reduction, for example via proper lock-in detection.

Next, the biodistribution of the magnetic labels was examined *ex vivo*. The animals were euthanized and different organs were excised, thoroughly washed from traces of blood and then analyzed via MPQ detector. As summarized in Table S2 and Fig. S3, in this experiment, the strongest MPQ signal was detected in the lungs, while significantly weaker signals were detected in the liver and spleen. The signals from other organs, including the brain, were negligible. These data are in agreement with the magnetic resonance imaging (MRI) of the whole animal performed separately, prior to organ extraction. Indeed, T2-weighted images (Fig.5) demonstrate a high concentration of disks in the lungs, liver and spleen with non-detectable uptake by other organs.

Finally, we studied the feasibility of MPQ-based detection of the actuation and manipulation of the disks directly in animal tissues. As we showed earlier, unlike conventional isotropic nanoparticles, the disks feature an interesting angular dependence of their MPQ-signal. Therefore, if one could magnetize the disks in tissue samples so that they physically rotate, the MPQ-signal should change. The disks rotational behavior (spontaneous or driven) depends on the characteristics of surrounding environment, such as its density, stiffness and viscosity, e.g., it may differ for different organ tissues.

This is what was observed in the experiment. Figure S4 shows that upon magnetizing the tissue samples (liver, spleen and lungs) by applying static magnetic field oriented along the MPQ-coil axis (along the detecting field) the signal gradually changes. Interestingly, different tissues types exhibit different relative magnitudes of the signal change, namely, the signal increases in both the liver (59% rise) and spleen (35% rise), while the signals in the lungs does not change. When the samples are magnetized with the field perpendicular to the detection coil field orientation, the signal decreases for the liver (~8%) and spleen (~12%) and, again, remains unchanged for the lungs. Further magnetization of the same samples along the second orientation perpendicular to that of the detection coil further decreases the signals of liver (12% total decrease) and spleen (23% total decrease) samples. Although explaining these data is not trivial, one could propose that some combination of MPQ/MPI techniques with disk-based labels can provide unique insight on local micro-rheology of the tissues or the cellular environment *in vivo*. For instance, while many studies have shown that stiffness of individual diseased (cancerous) cells can differ from healthy ones, performing such analysis *in vivo* has always been a challenge.

**Conclusions**

The impact of modern nanomaterials and engineered architectures on biological modulation, [40] bioanalytical techniques, [41-43] and biomedical technologies [44] can hardly be overestimated. Magnetic nanostructures engineered using top-down techniques have been exploited in the burgeoning area of biosensors as multi-spectral MRI contrast enhancement labels,[45-47] synthetic antiferromagnetic particles [48-49] and magneto-resistive sensors [50] for the detection of ultralow concentrations of molecular markers. Here we demonstrate that microfabricated magnetic vortex nanomaterials can be used as ultrasensitive, tunable, nonlinear magnetization labels. The magnetically soft, disk-shaped particles reveal a strong nonlinearity of their magnetization process due to irreversible transitions from the spin vortex to single-domain configuration. This feature of permalloy magnetic labels enables a sensitivity of MPQ on the level of $3.5 \times 10^{-9}$ emu or 39 pg of magnetic material. Moreover, detection of these materials was demonstrated in complex biological environments, including *ex-vivo* isolated organs and *in-vivo* animal models. Further optimization can be achieved by using a magnetic material with a higher saturation magnetization value, and tailoring the magnetocrystalline anisotropy. [51] Furthermore, varying geometric parameters of the disks, or introducing nanoscale interfacial and geometrical alterations (*e.g.,* exchange bias, [52-54] roughness, [55] or nanopatterning [56]) allows for controlling the vortex nucleation and annihilation, as well as vortex-core pinning and, thus, enhancing the *ac* nonlinearity effects in small fields. It is noteworthy that desired geometries and magnetic properties of the alloy disks are readily tunable and tightly controllable via micro- and nanofabrication methods. These labels, while being magnetically silent in the absence of external magnetic fields, possess unique and distinct size-, geometry- and material-dependent properties that can be remotely interrogated with high fidelity by means of low frequency circuitry. Furthermore, intensive translational studies, including tuning the properties of disk surfaces for enhancement of their *in vivo* stability and biocompatibility, blood circulation life-time, biosafety, side-effects, and long-term effects as well as affinity to sites of disease are required and can be achieved *via* biochemical functionalization. Beside the variety of detection and imaging needs in bioengineering and medicine, another obvious and potent area of application is magnetic tagging for biopharmaceuticals anti-counterfeit measures.

## Methods

**Fabrication of the disks**

Disk-shaped Permalloy ($Fe_{20}/Ni_{80}$ alloy) magnetic dots were prepared by optical lithography and ebeam evaporation using GaAs (for MPQ characterization of patterned dot arrays) and Si (for *in vivo* experiments) wafers as described elsewhere [3-4, 12, 57] with some modifications, as schematically depicted in Fig. S5. GaAs was chosen as the substrate since it can be cleaved readily into small sample pieces without the use of a dicing saw. Unlike previous reports on unbound disks, [12, 57] here a positive tone resist was used that allowed for patterning the disks directly on the substrate. This kind of sample is favorable for 2D magnetic detection model experiments. Samples with fixed disk diameter $D = 1.5$ μm, arranged on a square lattice with the period a = 3 μm, and thickness $h$ varied from 10 to 40 nm, were fabricated on 2 inches wafers and then sectioned into 3x3 mm² slides for further characterization. For the animal experiments the disks (diameter 1.5 μm) were lifted off the wafer and re-suspended in phosphate buffered saline (PBS) resulting in a concentration of ~$2 \times 10^9$ disks per 1 ml, following the procedures reported elsewhere.[12]

**Disks characterization**

The scanning electron micrographs of the samples were obtained using Raith eLINE Plus tool (InLens detector, 10 keV / 30-μm aperture). Magnetic properties of the samples were characterized using vibrating sample magnetometer (VSM; Model 7400, Lake Shore) and SQUID magnetometer (model MPMS-XL7, Quantum Design), as well as two MPQ-1[26-27] and MPQ-2 set-ups.[21, 25-28, 58]

**Animals.** BALB/c were purchased from Pushchino breeding facility, BIBC RAS, Pushchino, Russia. Anesthetization of the animals was performed with mixture of Zoletil (Virbac, France) and Rometar (Bioveta, Czech Republic): Tiletamine-HCl / Zolazepam-HCl / Xylazine-HCl in a dose 20/20/1.6 mg/kg. All procedures were approved by the Institutional Animal Care and Use Committee.

**MPQ detection of the magnetic disks** *in vivo*

Hundred microliter sample aliquot of the magnetic disks suspension in PBS was injected into an anesthetized animal's blood flow retro-orbitally. For blood circulation dynamics, the animal's tail was placed inside of the MPQ pick-up coil and gently fixed to restrict movement. For biodistribution and tissue magnetization studies, the organs were extracted, washed with PBS and divided into several pieces. Each piece was inserted inside the coil of the MPQ device for measurement. Magnetization of the tissue samples was performed by placing the samples in correct orientation between two brick magnets for 20 min.

**Magnetic resonance imaging (MRI)**

MRI was performed with an ICON 1T MRI system (Bruker, USA) using a mouse whole body volume *rf* coil. A two-dimensional gradient echo FLASH sequence was used (TR/TE=1000/5 ms, FA = 60°, resolution of 300 µm, FOV=80x50 mm, 1 signal average, TA = 2 min 47 sec; 17 slices per scan, slice thickness of 1 mm).

**Data availability**

The authors declare that all relevant data are available on request.

**Acknowledgements**

Work at Argonne, including use of the Center for Nanoscale Materials was supported by the U.S. Department of Energy, Office of Science, Office of Basic Energy Sciences, under contract no. DE-AC02- 06CH11357. This work was supported, in part, by the grants of Russian Science Foundation No 16-12-10543 (development of ultra-sensitive MPQ detectors) and No 16-19-00131 (AFM investigations of nanostructures and their detectability studies for prospects of biomedical applications The support of the Ministry of Education and Science of the Russian Federation in the framework of Increase Competitiveness Program of NUST «MISiS» (№K3-2017-069), implemented by a governmental decree dated 16th of March 2013, N 211 is gratefully acknowledged).

**Additional Information**

The authors declare no competing financial interests. Supplementary information accompanies this article.

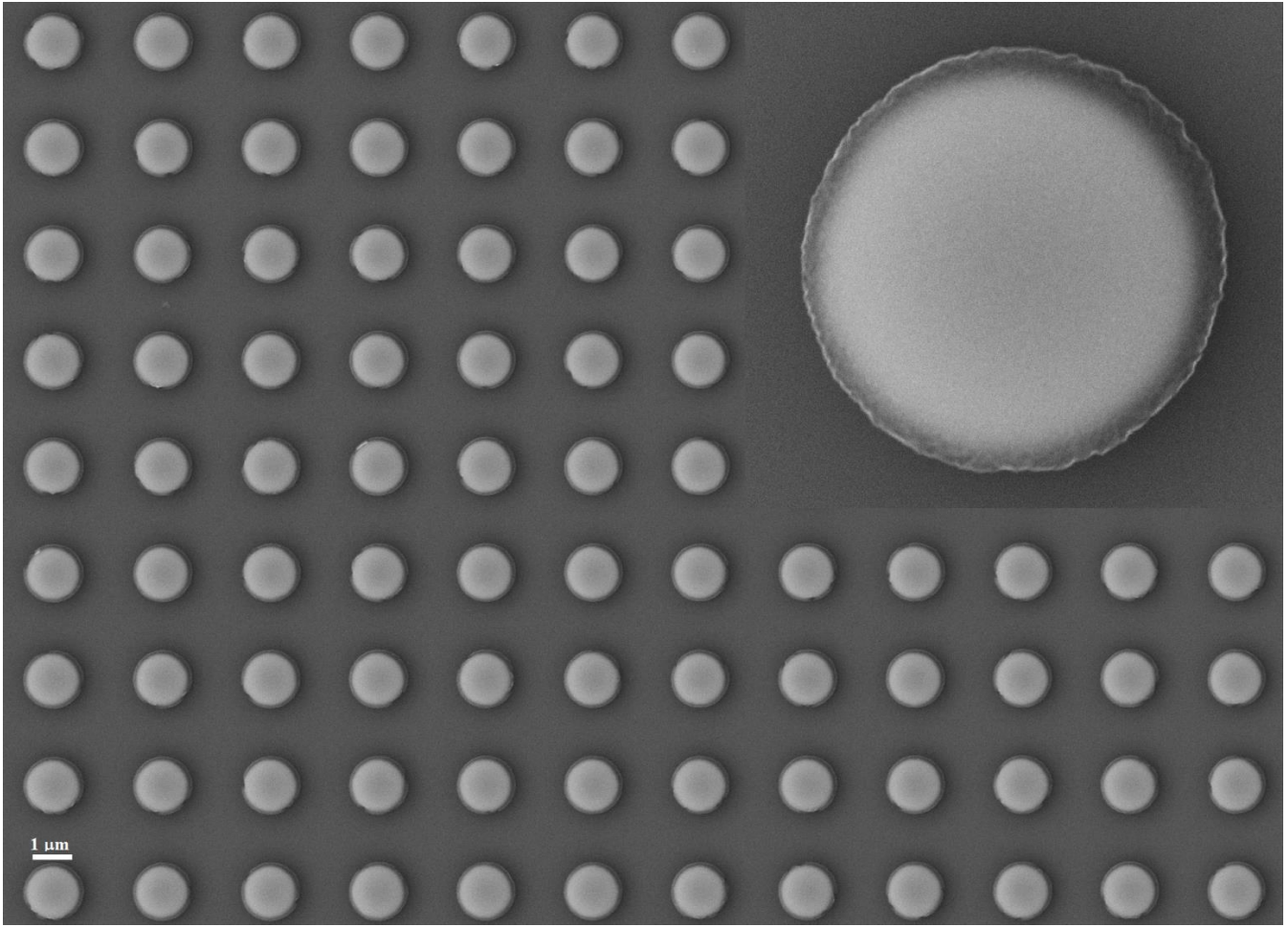

Figure 1. Representative scanning electron microscope image of 20-nm thick magnetic disks prepared by optical lithography and magnetron sputtering. The 1.5-μm diameter disks are arranged on a square lattice with a period of 3 μm.

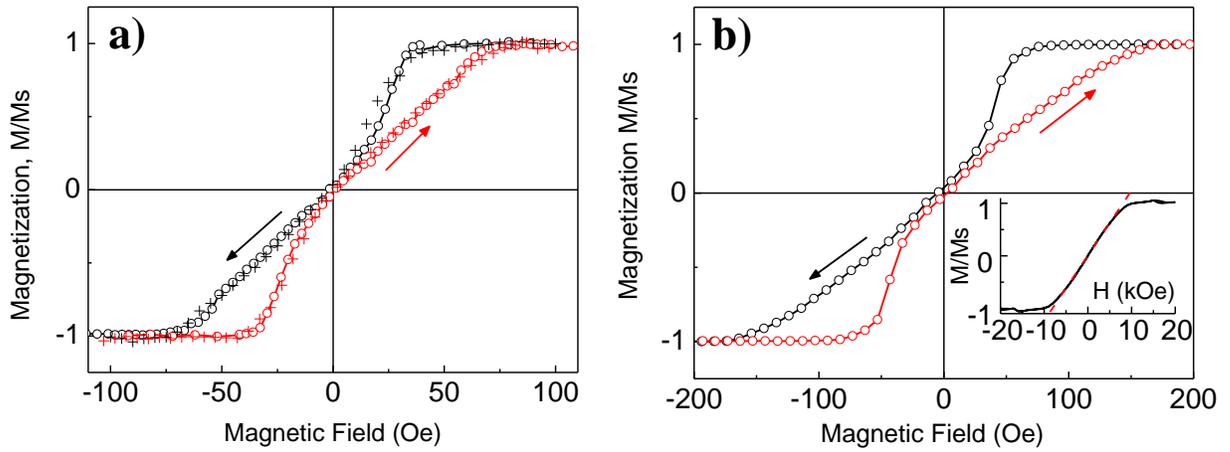

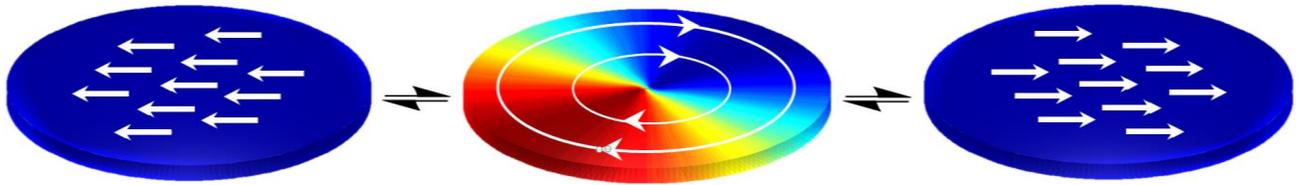

Figure 2. Normalized hysteresis loops for arrays of permalloy microdisks of 1.5-µm diameter measured by SQUID (circles) and VSM (crosses) in a magnetic field $H$ parallel to the disk plane for thickness $h = 20$ nm (a), and 40 nm (b). Normalized magnetization curve $M(H)$ with no hysteresis in the magnetic field perpendicular to the disk plane is shown in the insert (the diamagnetic background of the GaAs substrate is subtracted). (c) In remanence the samples possess no net magnetic moment, which is indicative of the spin vortex state (central image). In the saturated state the vortices annihilate and the sample becomes uniformly magnetized (left, and right images for negative, and positive magnetic fields, correspondingly).

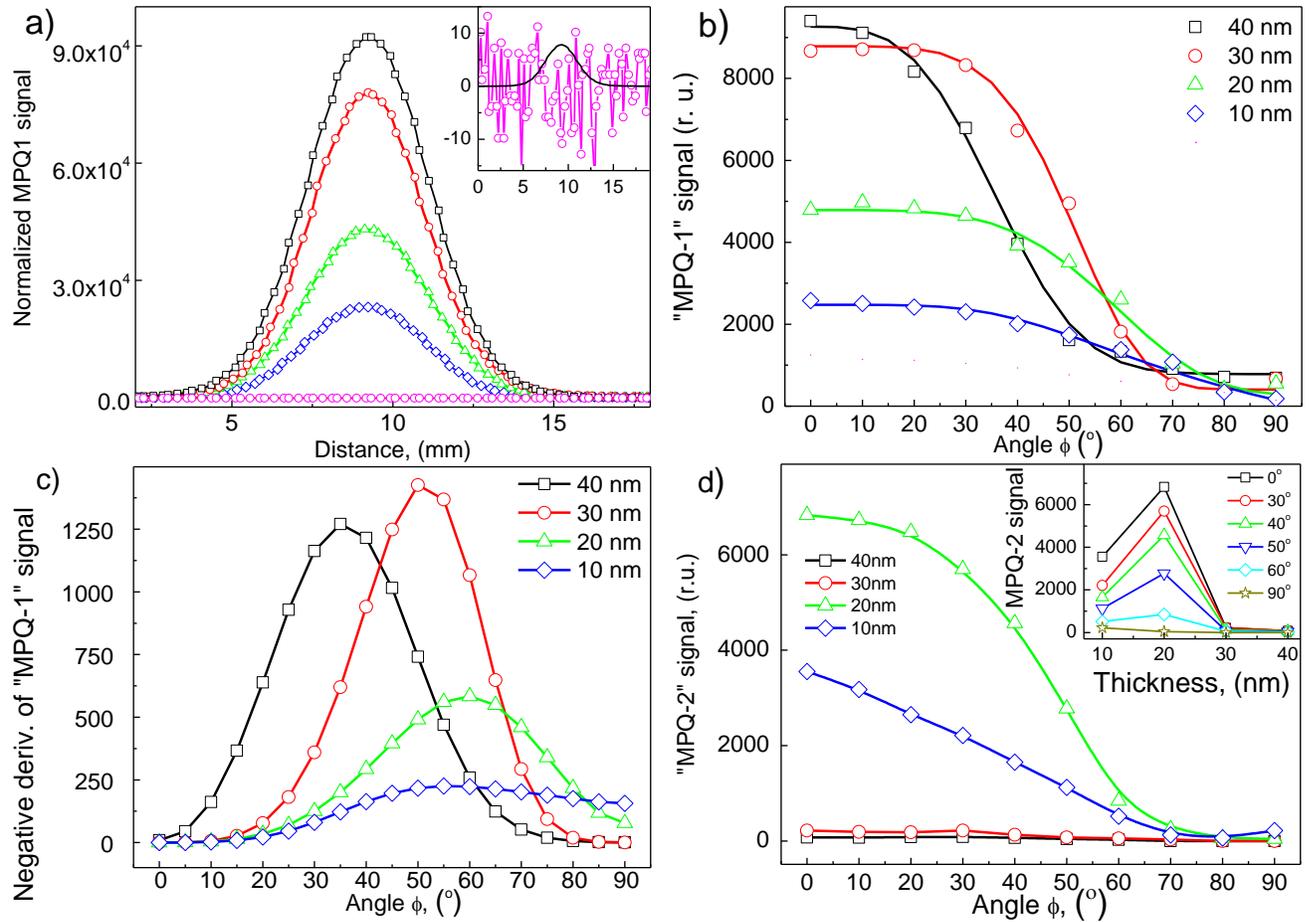

Figure 3. (a) The dependence of MPQ-1 sensor signal (in relative units) *vs*. slide position for 3 x 3 mm$^2$ size slide with the array of the 40 nm (squares), 30 nm (red circles), 20 nm (triangles), 10 nm (diamonds) thick magnetic disks, and for a bare wafer of the same size (magenta circles). Inset: the same dependences for the signal of 30-nm thick disks divided by a factor of 10$^4$ (black line), and the noise signal for the unloaded slide (open circles) (b) Angular dependences of MPQ-1 sensor signal normalized for 1 mm$^2$ of the array of disks of thicknesses of 40 (squares), 30 (circles), 20 (triangles), and 10 nm (diamonds). Zero degrees corresponds to the disk orientation being parallel to the field of the detection coil (c) Normalized negative derivative of the angular dependences of MPQ-1 sensor signal for disks of thicknesses of 40 (squares), 30 (circles), 20 (triangles), and 10 nm (diamonds). (d) Angular dependences of lower field MPQ-2 sensor signal $S(h, \varphi)$ normalized for 1 mm$^2$ square of the microdisk array of thickness 40 (squares), 30 (circles), 20 (triangles), and 10 nm (diamonds). Inset is the dependence of the signals on the thickness of the discs for different angles $\varphi$.

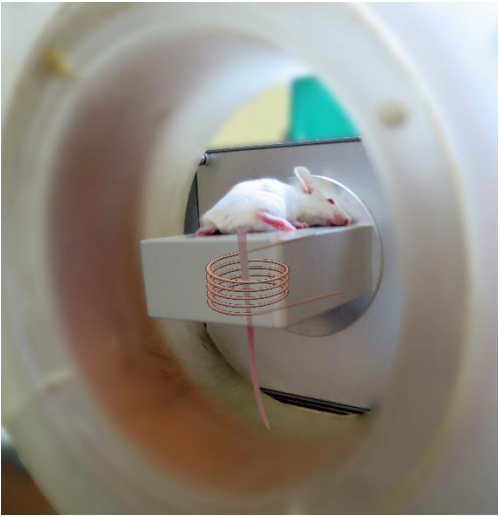
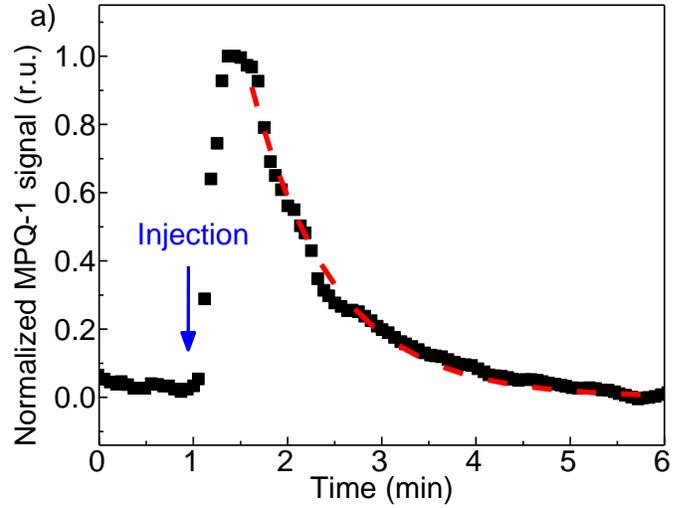
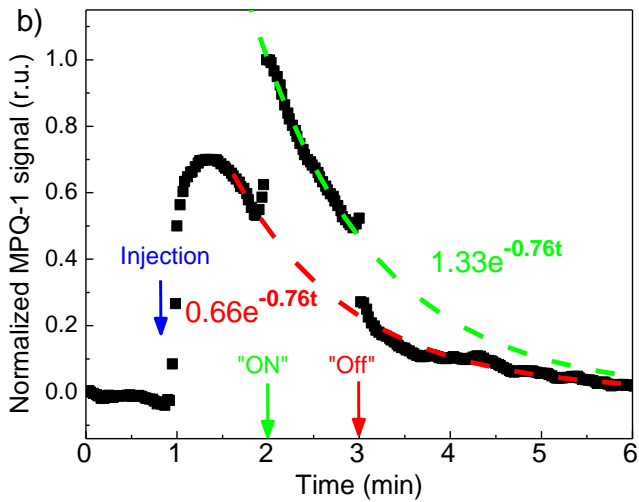
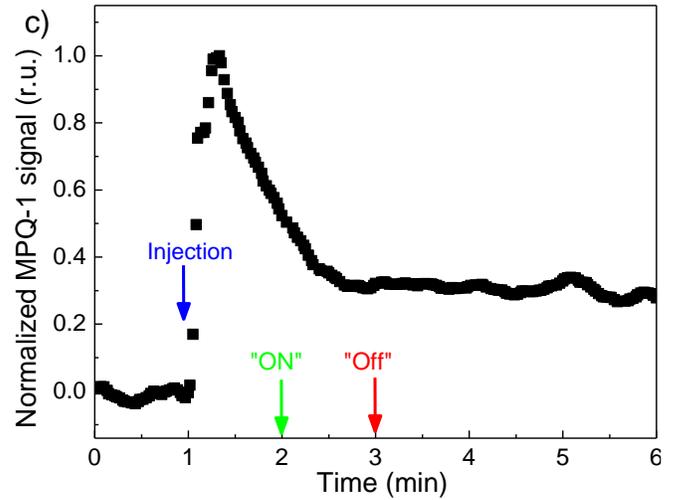

Figure 4. Experimental setup for the study of blood-circulation kinetics of the magnetic disks *in vivo*. The tail of the anesthetized animal is passed through the measurement coil of the MPQ device while the whole animal is placed within the large coil, which generates a *dc* magnetic field when turned on. The disks are injected retro-orbitally and detected in real time as they pass through the veins and arteries of the animal's tail (upper left image). (a) Representative blood circulation kinetics of the magnetic disks in mice: injection, distribution within the blood volume and clearance. (b) Modulation of the MPQ response of the magnetic disks by the external *dc* field in blood flow *in vivo*. (c) The control experiment showing absence of the signal modulation for the conventional (non-vortex) magnetic particles.

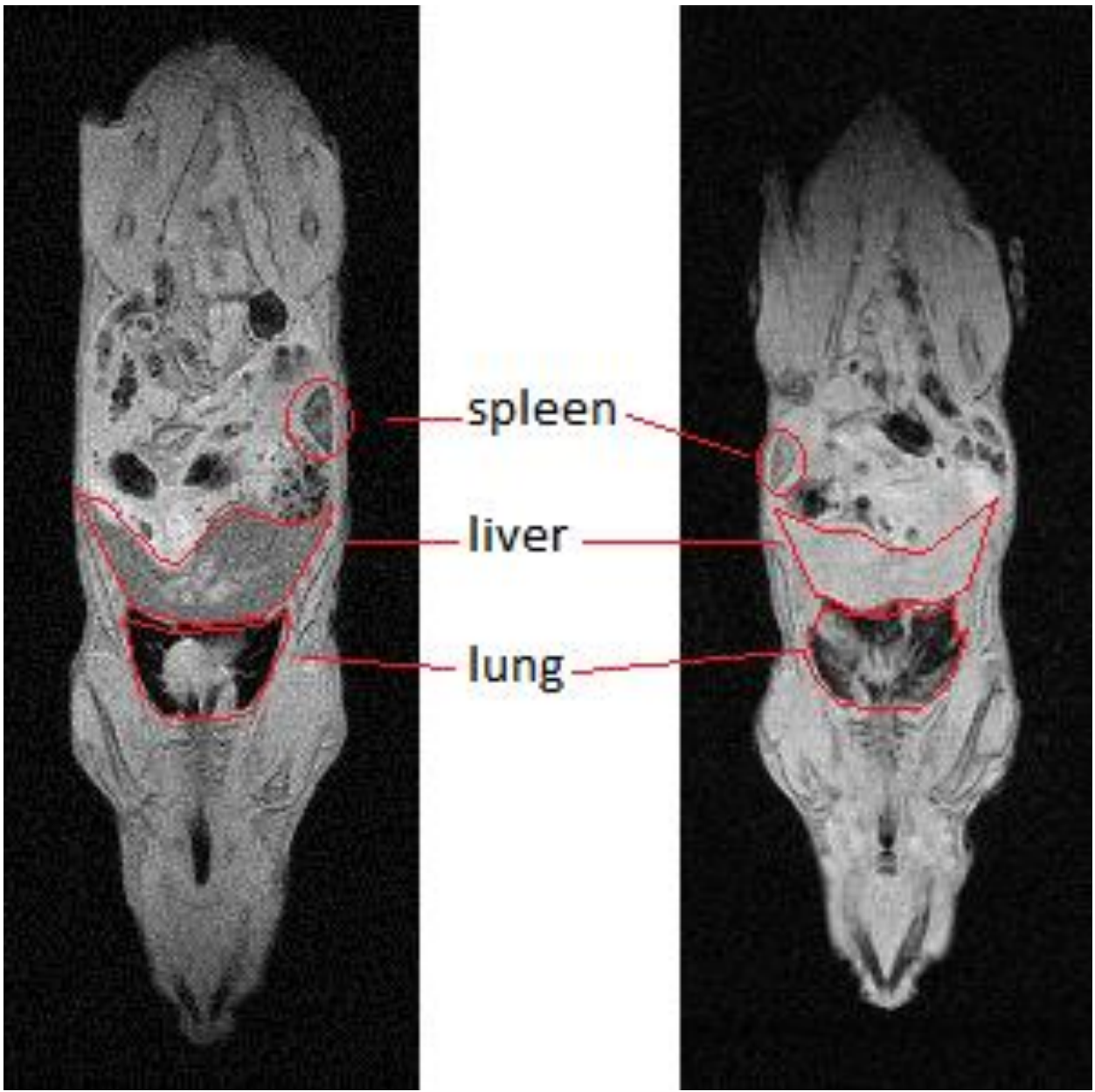

Figure 5. MRI visualization of distribution of the magnetic disks in the animal. Contrasting of the image for the animal injected with the disks (left) as compared to the control mouse without disk injection (right) can be seen in the liver, spleen and lungs regions as darker color indicating the presence of the disks.

**Table of Contents Graphic**

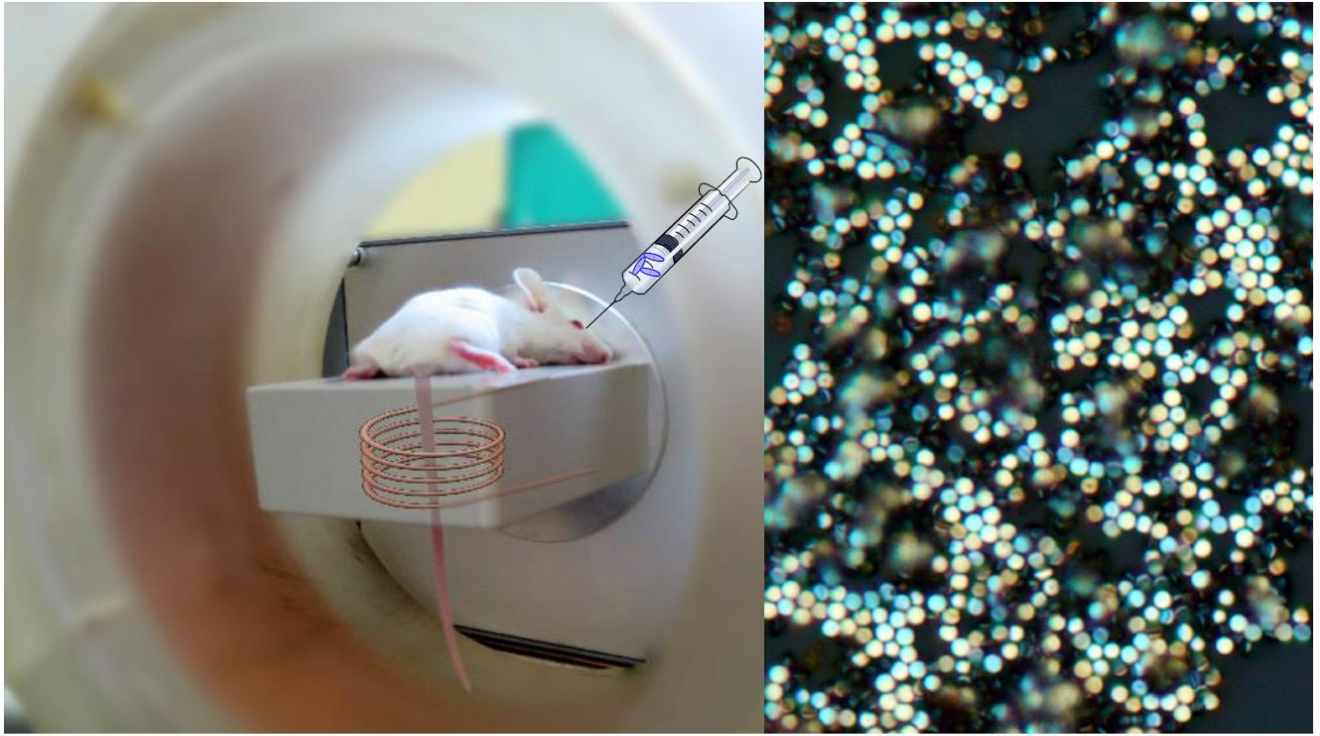

M. Nikitin *et al.*,

**Ultrasensitive Detection Enabled by Nonlinear Magnetization of Nanomagnetic Labels**

**Supplementary Information.**

**Ultrasensitive Detection Enabled by Nonlinear Magnetization of Nanomagnetic Labels**

M. P. Nikitin[1,2,3], A. V. Orlov[3], I. L. Sokolov[2], A. A. Minakov[3], P. I. Nikitin[3], J. Ding[1], S. D. Bader[1], E. A. Rozhkova[4], V. Novosad[1,5].

[1] Materials Science Division, Argonne National Laboratory, Argonne, IL 60439, USA.

[2] Moscow Institute of Physics and Technology, Dolgoprudny 141700, Russia.

[3] Prokhorov General Physics Institute, Moscow 119991, Russia.

[4] Center for Nanoscale Materials, Argonne National Laboratory, Argonne, IL 60439, USA.

[5] National University of Science and Technology (MISiS), Moscow, 119049, Russia.

Content:

I. Figure S1
II. Analysis of hysteresis cycles in alternating magnetic fields; Figure S2, and Table S1.
III. Figure S3
IV. Table S2
V. Figure S4
VI. Figure S5

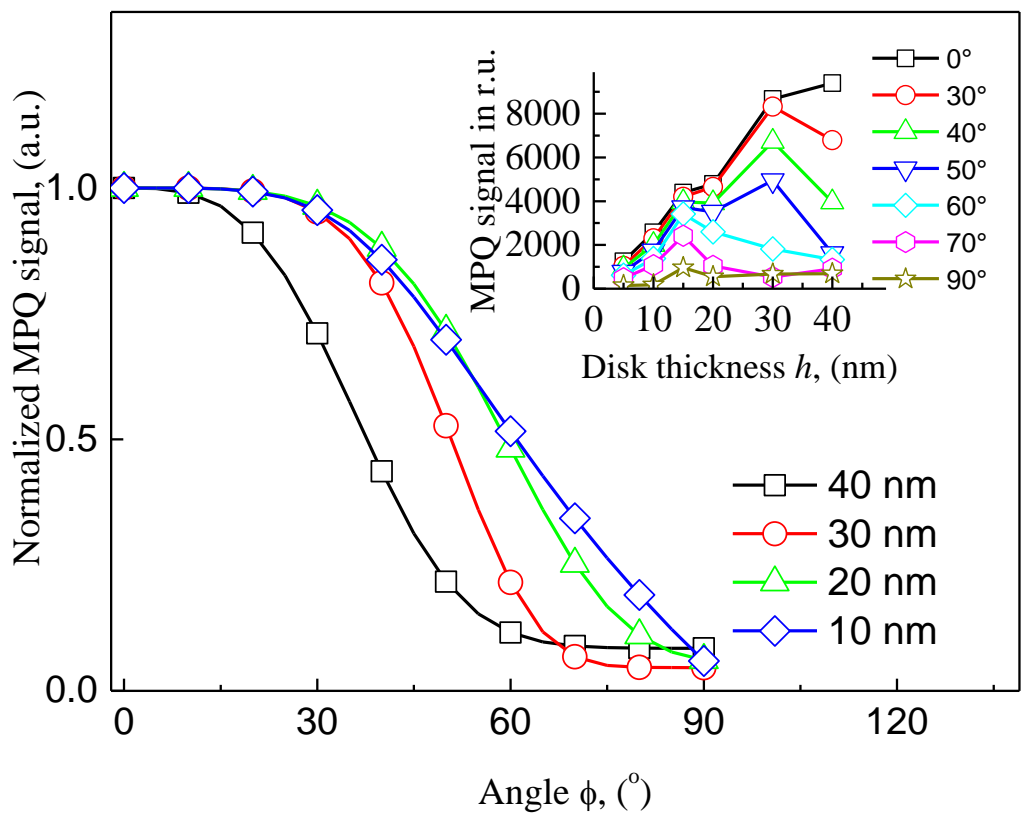

Fig. S1: Angular dependences of *normalized* MPQ sensor signal $S(h, \varphi)$ for the same parameters as in Fig. 4 at thicknesses $h = 40$ nm (squares), 30 nm (circles), 20 nm (up triangle), and 10nm (diamonds). The dependence of $S(h, \varphi)$ *vs*. $h$ at $\varphi = 0°$ (squares), 30° (circles), 40° (up triangle), 50° (down triangle), 60° (diamonds), 70° (hexagons), 90° (stars) are shown in the insert.

III. **Analysis of hysteresis cycles in alternating magnetic fields**

The magnetization curves (see Fig. 2) can be described by descending $\mathsf{M}_d(H_d)$ and ascending $\mathsf{M}_a(H_a)$ branches of the hysteresis loops for descending $H_d$ and $H_a$ ascending magnetic fields. According to the Weierstrass approximation theorem the descending and ascending branches can be approximated by polynomial functions:

$$\mathsf{M}_d(H_d) = B_{0d} + B_{1d}H_d + B_{2d}H_d^2 + B_{3d}H_d^3 + B_{4d}H_d^4 + \ldots, \tag{s1}$$

$$\mathsf{M}_a(H_a) = B_{0a} + B_{1a}H_a + B_{2a}H_a^2 + B_{3a}H_a^3 + B_{4a}H_a^4 + \ldots. \tag{s2}$$

A few major coefficients of Eq. (s1) and Eq. (s2) determined with a Mathcad program for normalized hysteresis loops are listed in Table S1.

**Table S1** Primary coefficients of descending and ascending branches of the polynomial functions approximating the normalized hysteresis loop for nanodisks of 1.5 μm diameter and thickness $h$.

| Thickness $h$, in nm | Branch | $10^5 \cdot B_1$, dimensionless | $10^5 \cdot B_2$, Dimensionless | $10^5 \cdot B_3$, dimensionless | $10^5 \cdot B_3/B_1$, dimensionless |
|---|---|---|---|---|---|
| 10 | Descending | 4499.4 | 91.15 | 0.7326 | 16.3 |
| 10 | Ascending | 4259.1 | -146.0 | 2.2940 | 53.9 |
| 20 | Descending | 1463.8 | 24.73 | 1.1673 | 79.7 |
| 20 | Ascending | 1646.3 | -33.24 | 0.8495 | 51.6 |
| 40 | Descending | 990.6 | 13.47 | 0.0839 | 8.5 |
| 40 | Ascending | 924.1 | -12.38 | 0.1158 | 12.5 |

The result of the approximation of the hysteresis loops by the polynomial functions of thirteenth degree is shown in Fig. S2. The nonlinear terms of the 2$^{nd}$ and the 3$^{rd}$ degree in the descending and ascending branches of the hysteresis loops of the measured nanodisks are significant, which is essential for magnetic particle detection at combination frequencies of a double-component applied magnetic field. The signal at combination frequencies of the applied magnetic field arises due to the strong nonlinearity of the hysteresis loops of the circular magnetized nanodisks (see Fig. S32.

The polynomial functions $M_d(H_d)$ and $M_a(H_a)$ can be further utilized for estimation of nonlinear response of the disks for a double-component alternating magnetic field:

$$H = H_1 \cdot \cos(2\pi f_1 \cdot t) + H_2 \cdot \cos(2\pi f_2 \cdot t), \qquad (s3)$$

where $H_1$, $H_2$, and $f_1$, $f_2$, are the amplitudes and the frequencies of the components of the applied magnetic field (the components are collinear), respectively. The model calculations were performed at frequencies $f_1 = 150$ Hz, $f_2 = 150$ kHz and amplitudes $H_1 = 75$ Oe, $H_2 = 5$ Oe for nanodisks of thickness $h = 20$ nm, as well as at $H_1 = 150$ Oe, $H_2 = 10$ Oe for $h = 40$ nm.

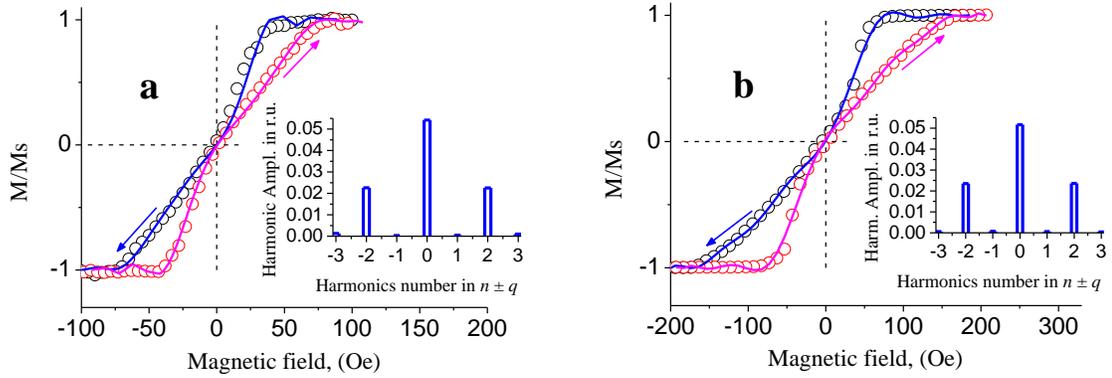

Fig. S2: The experimental hysteresis loops, the same as in Fig. 2 (circles), and polynomial functions (see Eq. (s1) and Eq. (s2)) approximating descending and ascending branches (solid lines) for the disk arrays of thickness $h = 20$ nm (**a**) and 40 nm (**b**). Amplitudes $A_p$ of harmonic oscillations of the magnetic moment in double-component alternating magnetic field at combination frequencies $f_p = (n \pm q) \cdot f_1$ for $q = 0, 1, 2, 3$ are shown in the inset.

The amplitudes $A_p$ of harmonic oscillations of the magnetic moment $M(H)$ at frequencies $p \cdot f_1$ in the double-component alternating magnetic field $H$ can be calculated for the following function approximating the hysteresis loop:

$$M(H) = \{M_d(H) \text{ if } \langle dH/dt \rangle_{AV} < 0\} \cup \{M_a(H) \text{ if } \langle dH/dt \rangle_{AV} > 0\}, \qquad (s4)$$

where $\langle dH/dt \rangle_{AV}$ is an average over the time interval $1/f_2$.

The amplitudes $A_p$ averaged over the time interval $1/f_1$ (as measured in the experiment) were calculated as follows:

$$S_p = 2f_1 \int_0^{1/f_1} \mathsf{M}(H) \cdot \sin(2\pi p \cdot f_1 \cdot t) dt, \qquad (s5)$$

$$C_p = 2f_1 \int_0^{1/f_1} \mathsf{M}(H) \cdot \cos(2\pi p \cdot f_1 \cdot t) dt, \qquad (s6)$$

and $A_p = \sqrt{S_p^2 + C_p^2}$.

The model calculations were performed for $f_2 = n \cdot f_1$ at $n = 1000$. The harmonic amplitudes $A_p$ at combination frequencies $f_p = (n \pm q) \cdot f_1$ for $q = 0, 1, 2, 3$ are shown in Fig. S2. Strong satellites of the driving harmonic $H_2 \cdot \cos(2\pi f_2 \cdot t)$ at frequencies $f_2 \pm 2l \cdot f_1$ (for $l = 1, 2, 3,...$) arise due to the strong nonlinearity of the hysteresis-loop $\mathsf{M}(H)$. No satellites occur at frequencies $f_2 \pm (2l + 1) \cdot f_1$ because of odd symmetry of the complete cycle of the hysteresis-loop $\mathsf{M}(H)$. Fortunately, the detection and the filtering of the signal at frequencies $f_2 \pm 2l \cdot f_1$ are easier to perform when no satellites occur at frequencies $f_2 \pm (2l + 1) \cdot f_1$. Normally the signal at frequencies $f_2 \pm 2 \cdot f_1$ is detected in MPQ sensors.

| Organs | MPQ signal, (arb units) | Disks weight normalized to organ mass (ng / mg) |
|---|---|---|
| Brain | 16 | 0.062 |
| Heart | 24 | 0.028 |
| Right kidney | 0 | 0.000 |
| Right thigh bone | 9 | 0.004 |
| Right thigh muscle | 15 | 3.48 |
| Skin | 7 | 0.056 |
| Spleen | 533 | 29.8 |
| Liver | 808 | 131 |
| Lungs | 4865 | 60.5 |

**Table S2:** MPQ-based detection of the disks in the organs of mice *ex vivo* after retro-orbital administration of the disks.

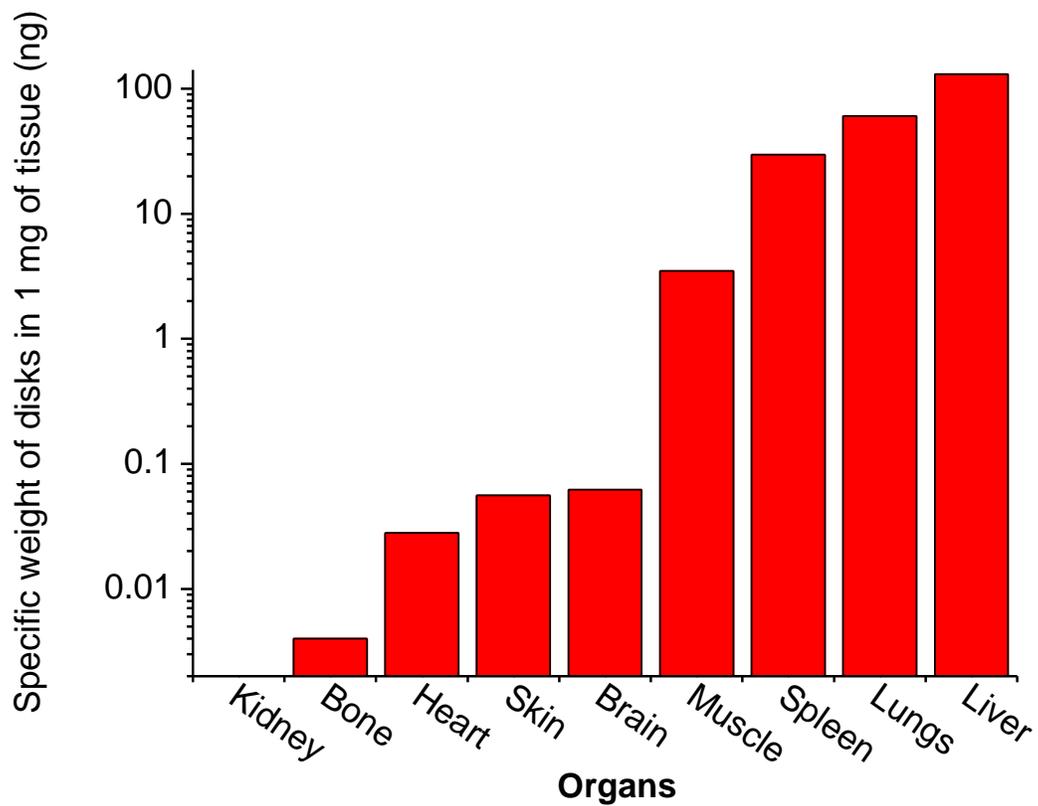

Figure S3 MPQ-based detection of the disks in the organs of mice *ex vivo* after retro-orbital administration of the disks (data taken from Table S1).

| | Parallel | Before magnetization | Perpendicular 1 | Perpendicular 1&2 |
|---|---|---|---|---|
| Field scheme | 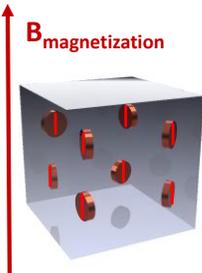 | 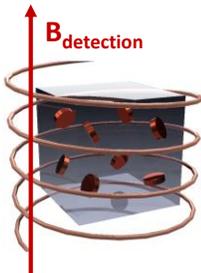 | 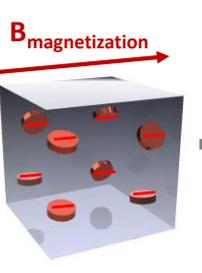 | 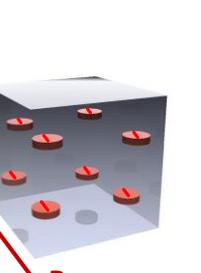 |
| Liver | 1,59 ± 0,18 | 1 | 0,92 ± 0,02 | 0,88 ± 0,03 |
| Spleen | 1,35 ± 0,16 | 1 | 0,88 ± 0,07 | 0,77 ± 0,21 |
| Lung | 0,98 ± 0,03 | 1 | 0,98 ± 0,02 | 0,98 ± 0,01 |

Figure S4. MPQ-based detection of the rotation of the disks within the organs (liver, spleen, lungs) of mice injected with magnetic disks. The table shows changes in the MPQ-signals of the small pieces of each organ (n=3 pieces) after their magnetization with the field oriented parallel to the field of the MPQ detection coil (20-min magnetization), in one perpendicular direction (20-min magnetization) or subsequently in one perpendicular orientation (20-min) and then in the other perpendicular orientation (another 20 min). The scheme illustrates that magnetization in both perpendicular directions is needed in order to orient all disks perpendicular to the detection coil's field, and consequently is much harder to realize than magnetization in the parallel orientation.

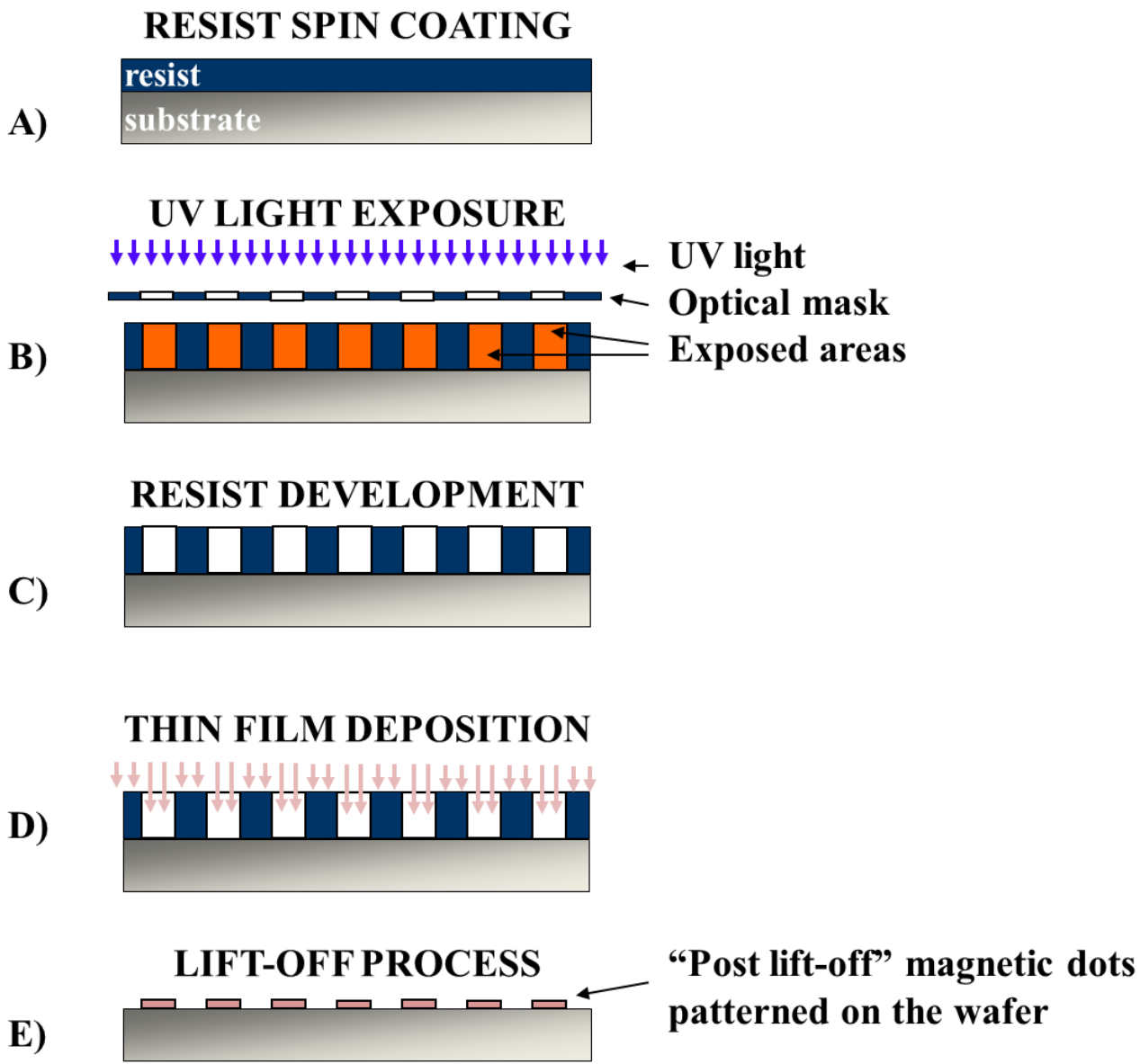

Figure S5. Scheme of samples preparation using optical lithography and magnetron sputtering. **A-E**. Patterned disks samples fabrication: the process starts with a positive tone photoresist spin coating on a silicon or GaAs wafer (**A**). An optical mask is placed in contact with the layer of pre-baked photoresist and illuminated with UV light (B). An organic solvent dissolves and removes photoresist that is not exposed (**C**). Magnetron sputtering is used to deposit thin layer of magnetic materaisl (D). After the lift –off process in organic solvent the patterned disk array is defined on the wafer (C). For biological experiments the negative tone resist is used instead, and, the disks are released from the wafer by lift-off process.

36